\documentclass[useAMS,usenatbib]{mn2e}
\usepackage{color}
\usepackage{amssymb}
\usepackage{amsmath}
\usepackage{txfonts}
\usepackage{graphicx}
\usepackage{natbib}
\usepackage{epsfig}
\usepackage{xspace}

\def\fermi{Fermi\xspace}
\def\hessj{HESS~J0632+057\xspace}

\begin{document}
\title[VHE \fermi/LAT detection of \hessj]{Very high energy Fermi/LAT detection of HESS J0632+057} 
\author[D. Malyshev, M.Chernyakova]{D.Malyshev$^{1}$, M.Chernyakova$^{2,3}$ \\ 
$^{1}$ INTEGRAL Science Data Center, Chemin d'Ecogia 16, 1290 Versoix, Switzerland\\
$^2$ Dublin City University, Dublin 9, Ireland\\
$^3$ Dublin Institute for Advanced Studies, 31 Fitzwilliam Place, Dublin 2, Ireland
}

\date{Received $<$date$>$  ; in original form  $<$date$>$ }
\pagerange{\pageref{firstpage}--\pageref{lastpage}} \pubyear{2014}

\maketitle
\label{firstpage}
\begin{abstract}
{ We report on the results of $\sim 7$~yrs of the very-high energy (10-600~GeV) observations of \hessj with \fermi/LAT. In the highest energy band, $200-600$~GeV, the source is clearly detected with the statistical significance $\gtrsim 3.6\sigma$ at orbital phases $0.2-0.4$ and $0.6-0.8$ at which \hessj is known to demonstrate enhanced emission in TeV energy band. The analysis did not reveal the emission from \hessj at lower energies and different orbital phases. Using the upper limits on source's flux we locate the break of the spectrum to $>80$~GeV and low-energy slope $<2.0$ ($2\sigma$ statistical significance).
}
\end{abstract}

\section{Introduction} 
\label{sec:intro}

\hessj is a rare representative of a TeV binary class.  All five known binaries of this class host either O-type main sequence star or a Be star with a circumstellar disk and belong to the broad high-mass X-ray binary (HMXB) population. 
Peculiarly the spectral energy distribution in these systems is similar and dominating by the emission at GeV-TeV energies. The nature of this peculiarity is not completely understood, but allow to suggest the similar for all systems geometry and the nature of compact object, see~\citet{dubus_review13} for the review. At least one of system, PSR B1259-63 is known to be powered by a young pulsar~\citep{Johnston1992} and for several others the same type of compact object was suggested~\citep{neronov08,torres10,zdz10,durant11,moritani15}.

Among the TeV binaries \hessj is the only source which was not detected at GeV energies so far, which may indicate the system somewhat differs from another TeV binaries. Discovered in 2007 during HESS observations of Monoceros region~\citep{hess_j0632} as an unidentified point source with the galactic coordinates ($\ell, b$) = (205.67 ; -1.44) it was considered to be TeV binary candidate due to its spatial coincidence with the Be star MWC~148~\citep{hess_j0632,hinton09}. Later, the system was detected in radio~\citep{skilton09}, soft X-rays~\citep{falcone10} and TeV (VERITAS and MAGIC~\citep{magic12,veritas13}) energy bands. The observations of \hessj by SWIFT taken between 2009 and 2011 allowed to determine the orbital parameters of the system ($P_{orb}= 320 \pm 5$~d ; $e=0.83 \pm 0.08$ ; $T_0$=MJD~54857.0 ; $\phi_0=0.967 \pm 0.008$), see~\citet{bongiorno11}. The distance to the source was estimated as $\sim 1.4$~kpc~\citep{casares12}.

During the orbital cycle \hessj demonstrates in X-rays two clear peaks of the emission at orbital phases $\sim 0.2-0.4$ and $\sim 0.6-0.8$~\citep{bongiorno11}. Recently, the similar structure of the orbital lightcurve in TeV energy range was reported by~\citet{veritas15}. 
In February 2011 during the orbital phase $\sim 0.3$ \hessj demonstrated an outburst, detected in X-rays~\citep{swift_atel} and TeV energies~\citep{veritas_atel,jogler11,maier11}. During this flare the spectrum of the system remained consistent with a powerlaw of index $\Gamma=2.5$  observed earlier by HESS and MAGIC along the orbit~\citep{hess_j0632,magic12} and the slope at phases $0.2-0.4$ and $0.6-0.8$, as seen by VERITAS~\citep{veritas15}. 

Despite the significant efforts, \hessj was not detected in GeV band~\citep[see e.g.][]{mori13,caliandro13,hill13} with the upper limits suggesting turnover in the spectrum at energies $>10$~GeV~\citep{caliandro13}. Particularly, the non-detection of the system can be explained by its location in a crowded Monoceros Loop region. For \fermi/LAT with $\gtrsim 1^\circ$ PSF at $1$~GeV energy, the emission in \hessj vicinity is strongly dominated by the bright $\gamma$-ray pulsar PSR~J0633+0632 (3FGL J0633.7+0632), located only $0.8^\circ$ away. 

In order to minimize the contamination by PSR~J0633+0632, the previous studies~\citep{mori13,caliandro13,hill13} used for the analysis the data taken during the ``off-pulse'' phases of this pulsar, reducing the total available exposure on \hessj by $\sim 40\%$

Below we present the results of the analysis of the very-high energy (VHE, $>10$~GeV) \fermi/LAT data on \hessj, focusing additionally on the orbital phases at which the source is bright in TeV energy band. Due to the presence of the high-energy cut-off in the spectrum of PSR~J0633+0632 at 3.2~GeV~\citep{3fgl}, the analysis region was not significantly contaminated by the presence of the pulsar and does not suffer from the loss of the statistic. Such an approach for the first time allowed the detection of \hessj with \fermi/LAT.

\section{\fermi/LAT data analysis} 
\label{sec:data}
The Large Area Telescope (LAT) is one of the instrument of \fermi satellite operating since 2008 at energies $\gtrsim 0.1$~GeV. It has wide field of view of $\sim 2.4$~sr at 1~GeV and observes the entire sky every 3~hours~\citep[see full details of the instrumentaion in][]{atwood09}

\begin{figure*}
\includegraphics[width=0.55\linewidth]{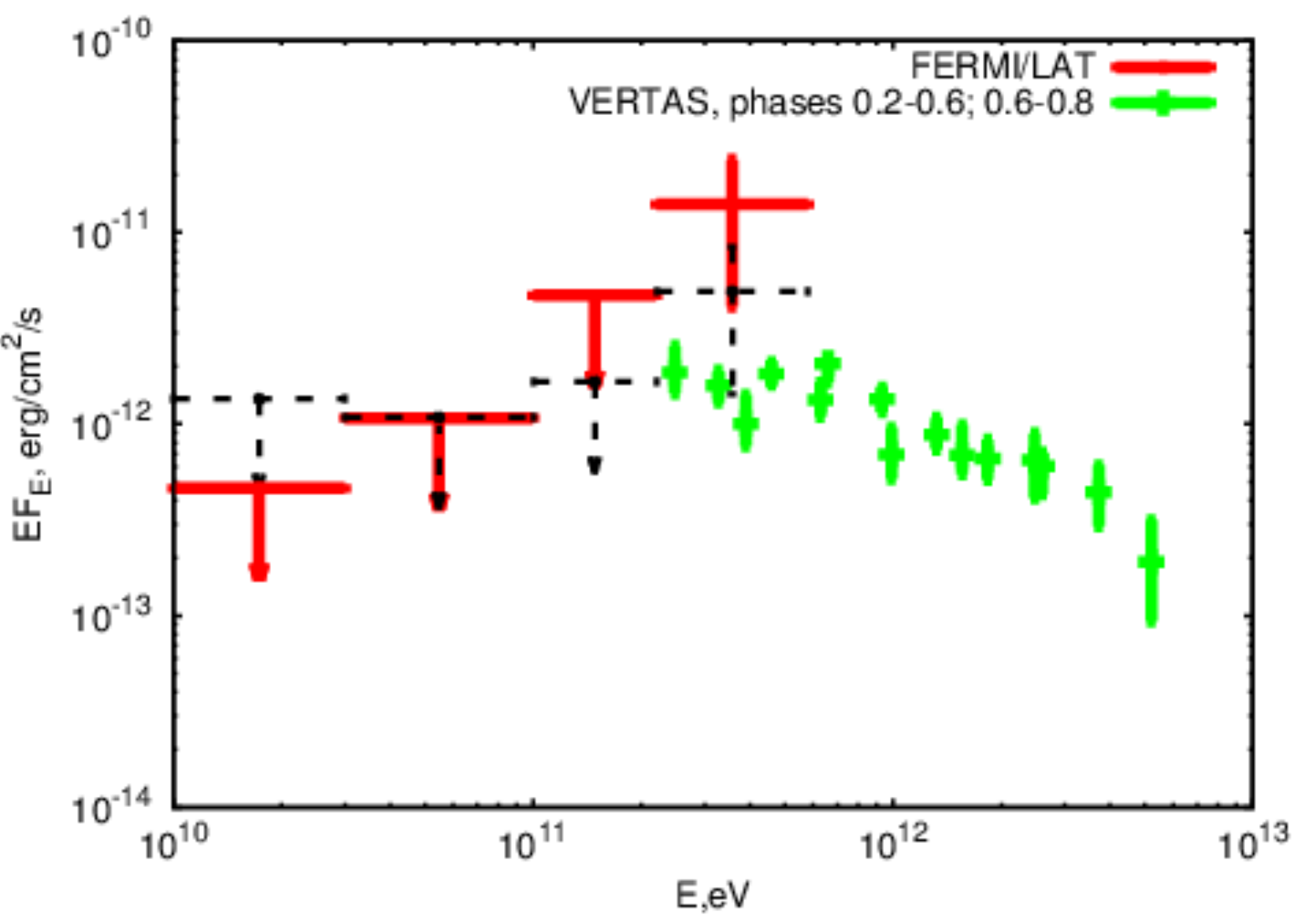}
\includegraphics[width=0.35\linewidth]{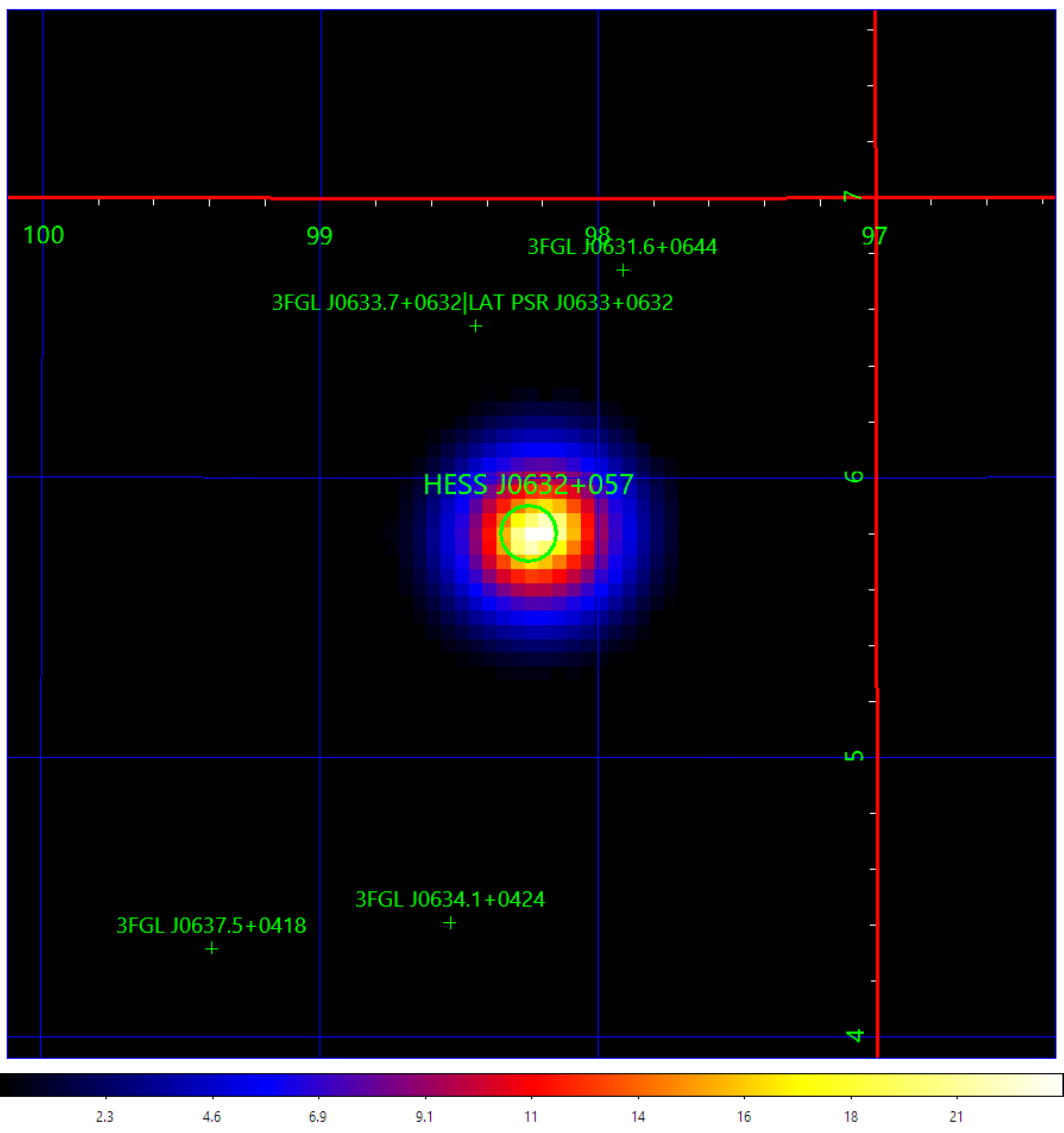}
\caption{Left:The spectrum of \hessj as seen by \fermi/LAT (solid red points) together with VERITAS spectrum~\citep{veritas15} of the source taken during the same orbital phases. The black dashed points correspond to the spectrum averaged over all orbital phases. Right: TS map of \hessj region. The positions of 3FGL catalogue sources are shown with green crosses. The significance of the source on the map is proportional to $\sqrt{TS}$}
\label{fig:spec_map}
\end{figure*}

In our analysis we consider all available \fermi/LAT data on \hessj taken during MJD,54682--57328 (2008 Aug.~04 -- 2015 Nov.~02). The data were processed with the standard LAT analysis software (\texttt{v10r0p5}) and analysed with \texttt{P8R2} response functions (\texttt{CLEAN} class photons). We have applied the standard cleaning to suppress the effect of the Earth albedo background, excluding time intervals with the Earth in the field of view (FoV; when its centre was $>52^\circ$ from the zenith), and those in which a part of the FoV was observed with the zenith angle $>90^\circ$.

Additionally we explicitly limit our analysis for energies $>10$~GeV in order to reduce the contamination on \hessj by the bright nearby pulsar LAT PSR~J0633+0632. For the region of $18^\circ$ radius around \hessj we perform binned likelihood analysis, based on a fitting of a model of the region to the data. The spatial model includes diffuse Galactic and extragalactic backgrounds (\texttt{gll\_iem\_v06.fits} and \texttt{iso\_P8R2\_CLEAN\_V6\_v06.txt} templates) and the sources from the  4~years (3FGL) \fermi catalog~\citep{3fgl}. We split the whole energy range into narrow energy bins, performing the fitting procedure in each bin separately. In each bin, the spectral shapes of all sources are assumed to have power law spectral shape with $\Gamma=2$. The spectral shapes of diffuse Galactic and extragalactic backgrounds are given by the corresponding templates. The normalisations of the fluxes of all sources and the diffuse background are treated as free parameters during the fitting. The analysis is performed with the python tools\footnote{fermi.gsfc.nasa.gov/ssc/data/analysis/scitools/python\_tutorial.html} provided by \fermi/LAT collaboration. The upper limits are calculated with the \texttt{UpperLimits} python module provided with the \fermi/LAT software and correspond to the 95 per cent ($\simeq 2.5\sigma$) false-chance probability.

The likelihood-ratio test statistic (TS) is employed to evaluate the significance of each model source. The TS value of the source S is defined to be
$$\mbox{TS}_S=-2(\log\mathcal{L}_0 - \log\mathcal{L}_{S})$$
where $\mathcal{L}$ is likelihood function maximized during the fitting of the model with and without corresponding source, see e.g.~\citet{mattox96}. The TS value for the source with the $N$-parametric spectral model follows $\chi^2$ distribution with N degrees of freedom~\citep{wilks38}, which allows to estimate the significance of added source. For the source modelled with the powerlaw spectrum (2 d.o.f.) this results in the significance $\simeq\sqrt{TS}$ with the precision better than 10\% for $TS>20$.

Following VERITAS detection of the outbursts from the source at orbital phases 0.2-0.4 and 0.6-0.8, we initially constrain the spectral analysis only to these parts of the orbit.
The spectrum of \hessj obtained in such a way is shown in Fig.~\ref{fig:spec_map}, left (red solid points). 
At the highest energy bin (200-600~GeV) the test-statistic value of the detection of point-like source at the position of \hessj is $\sim 24$, which implies $\sim 4\sigma$ detection significance. The observed in this band energy flux is $F_{0.2-0.4;0.6-0.8}=(1.4\pm 1.0)\cdot 10^{-11}$~erg/cm$^2$/s (assuming $\propto E^{-2}$ spectrum).

The averaged over all orbital phases spectrum is shown in Fig.~\ref{fig:spec_map} with the black dashed points. The non-detection of the source at orbital phases other than $0.2-0.4$ and $0.4-0.6$ together with the assumption of the constant over time flux implies in this case somewhat lower flux at 200-600~GeV energies $F_0=(0.5\pm 0.3)\cdot 10^{-11}$~erg/cm$^2$/s  and $TS\sim 20$

In order to further localise the source we build the TS map of the vicinity of \hessj. The Fig.~\ref{fig:spec_map}, right shows the significance $\sim \sqrt{TS}$ of adding a point source to the described region's model (with removed \hessj source) during the orbital phases $0.2-0.4$ and $0.4-0.6$. The position of the source was found to be consistent with the position of \hessj within the uncertainty found with \texttt{gtfindsrc} tool ($\sim 0.1^\circ$, similar to the \fermi/LAT PSF at energies $\gtrsim 100$~GeV).

\section{Discussion and conclusions}
\label{sec:discussion}

Despite high formal significance of the detection of \hessj by \fermi/LAT at VHE energies, the total amount of the detected photons is rather small. Namely, during the time periods which correspond to the increased TeV flux in VERITAS observations, \fermi detected 2 photons with energies $223$~GeV and $578$~GeV.

We verify, that the formal significance obtained by the \fermi software is consistent with the following conservative estimate. The considered region of $18^\circ$ contains in total 35 photons from which 2 (attributed to \hessj source) are within $0.1^\circ$ from \hessj position. The statistical false-chance of the later event is 
$3.67\cdot 10^{-4}$ or $\sim 3.6\sigma$.

The detected photons arrived in two consequent orbits (MJD~55404 (2010, Jul. 27) and MJD 55934 (2012, Jan. 27). Together with somewhat higher flux comparing to VERITAS~\citep{veritas15} $\sim 9$~yrs mean orbital phases 0.2-0.4 and 0.6-0.8 flux, see Fig.~\ref{fig:spec_map} this may indicate the detection of \hessj during its increased activity period, seen by MAGIC and VERITAS in Feb.~2011, \citet{magic12,veritas13}.

The poor photon statistics does not allow also to determine the slope $\Gamma = -2.5 - 4.5$ of the source's spectrum $E^{-\Gamma}$ in 200--600~GeV energy band . However, the absence of the photons at 10-200~GeV energies allow to put constraints on the position of the break in the spectrum and the corresponding low-energy slope. 

Despite the strong flaring of \hessj in TeV energy band the slope of the observed spectrum remained constant. In what follow we additionally assume, that the GeV slope of the spectrum also remains constant over all time intervals corresponding to orbital phases $0.2-0.4$ and $0.6-0.8$ used in further analysis.

\begin{figure}
\includegraphics[width=\linewidth]{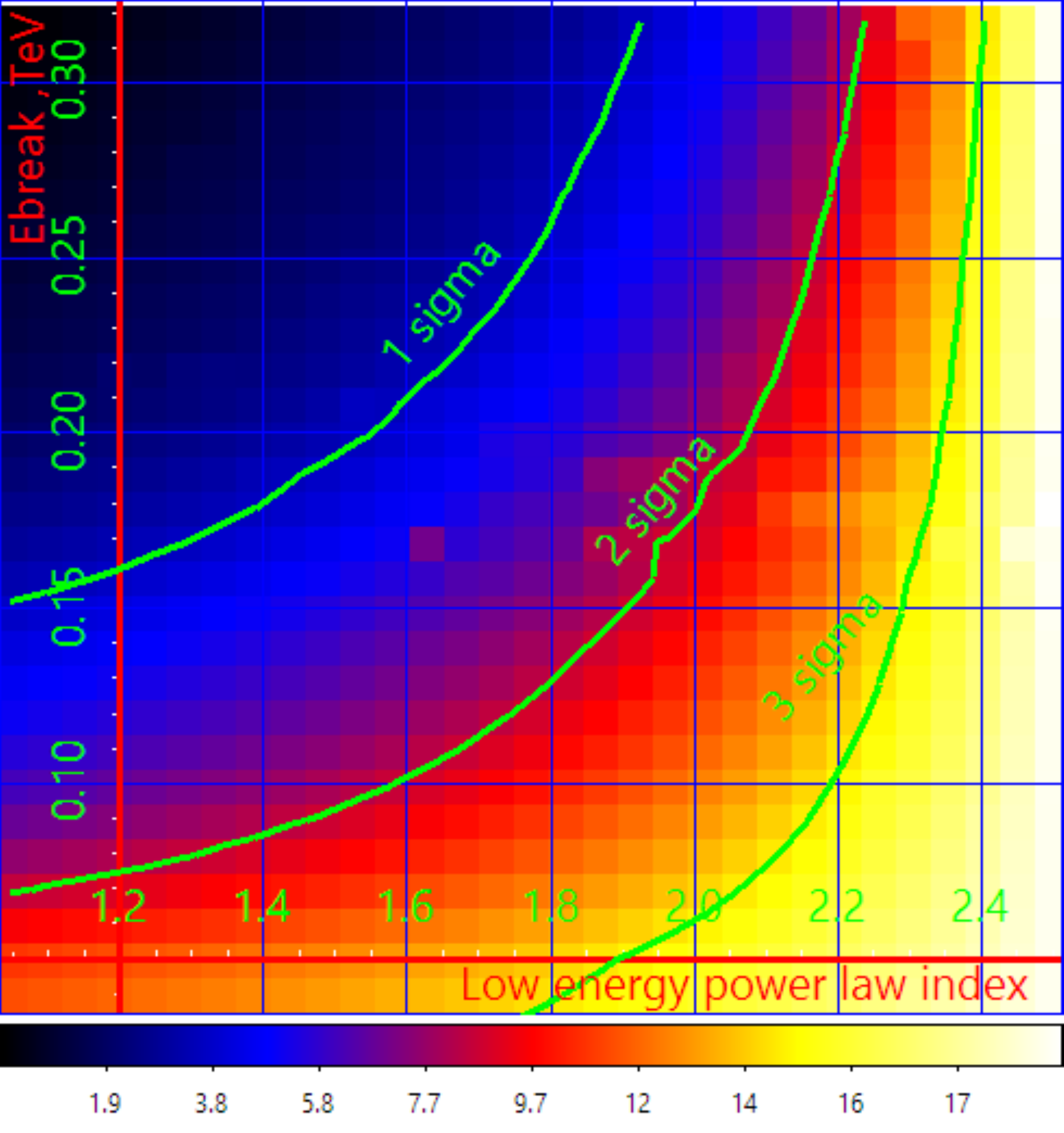}
\caption{Modelling \hessj spectrum with a broken power law with the corresponding low-energy slope and the position of the break we put the constrains on the parameters of the spectrum. The high-energy slope was fixed to be 2.5~\citep{veritas15}. Green contours shows $1\sigma$, $2\sigma$ and $3\sigma$ allowed parameter's range.}
\label{fig:broken}
\end{figure}

Fig.~\ref{fig:broken} shows the change of the best-fit log-likelihood of the model in which \hessj source was modelled with the broken powerlaw with corresponding low energy slope and the break position at $E_{break}$.  The slope at energies $E>E_{break}$ was fixed to best-fit value $\Gamma=2.5$ observed by HESS and VERITAS~\citep{hess_j0632,veritas15}. The flux of the source at~200~GeV was additionally constrained not to be lower than the flux observed by~\citet{veritas15}.

The change of the log-likelihood with adding N-parametric component to the model follows $\chi^2$ distribution with N degrees of freedom~\citep{wilks38}. 
Correspondingly, $1\sigma$, $2\sigma$ and $3\sigma$ allowed parameter's range for \hessj spectrum result from the change of the log-likelihood by $\sim 3.6$ ; $\sim 8.1$ and $\sim 14.2$ (3 added degrees of freedom) and are shown with green contours in Fig~\ref{fig:broken}. 

At $1\sigma$ ($2\sigma$) confidence the position of the break can be constrained as $E_{break}>150$~GeV  ($>80$~GeV). Together with the upper limit on the position of the break implied by VERITAS observations~\citep{veritas15} this allows to estimate the GeV energy slope of \hessj to be quite hard ($\Gamma_{1\sigma} < 1.5$ ; $\Gamma_{2\sigma} < 2.0$).

The obtained constraints on \hessj flux, the position of the break and 10-100~GeV slope of spectrum are consistent with the predictions from simple one-zone synchrotron self-Compton model proposed by~\citet{hinton09} and~\citet{skilton09}. In this model electrons injected with the spectrum $\sim E^{-2}$ are cooling in a magnetic $B=70$~mG and radiation ($U_{rad}=1$~erg/cm$^3$) fields. The ``striped pulsar wind model'' proposed by~\citet{petri11} to explain the emission from PSR B1259-63 could also explain the GeV-TeV observations and can be preferable taking into account recent optical observations suggesting pulsar nature of the compact object in this system~\citep{moritani15}.

Still, the quality of the analysed \fermi/LAT data does not allow to make firm conclusions on the nature of the emission from \hessj but shows the possibility of the \fermi and forthcoming $\gamma$-ray missions to detect this source at least during the increased flux state, similar to one, observed in 2010-12. We encourage a dedicated $\gamma$-ray campaign for observations of this source during the orbital phases $0.2-0.4$ and $0.6-0.8$ which together with TeV-range observations can clarify the behaviour of the source's spectrum at lower energies.

\noindent \textit{Acknowledgements.} 
This work was partially supported by the EU COST Action (COST-STSM-MP1304-28864) ``NewCompStar''. The authors thank SFI/HEA Irish Centre for High-End Computing (ICHEC) for the provision of computational facilities and support.

\def\aj{AJ}%
\def\actaa{Acta Astron.}%
\def\araa{ARA\&A}%
\def\apj{ApJ}%
\def\apjl{ApJ}%
\def\apjs{ApJS}%
\def\ao{Appl.~Opt.}%
\def\apss{Ap\&SS}%
\def\aap{A\&A}%
\def\aapr{A\&A~Rev.}%
\def\aaps{A\&AS}%
\def\azh{AZh}%
\def\baas{BAAS}%
\def\bac{Bull. astr. Inst. Czechosl.}%
\def\caa{Chinese Astron. Astrophys.}%
\def\cjaa{Chinese J. Astron. Astrophys.}%
\def\icarus{Icarus}%
\def\jcap{J. Cosmology Astropart. Phys.}%
\def\jrasc{JRASC}%
\def\mnras{MNRAS}%
\def\memras{MmRAS}%
\def\na{New A}%
\def\nar{New A Rev.}%
\def\pasa{PASA}%
\def\pra{Phys.~Rev.~A}%
\def\prb{Phys.~Rev.~B}%
\def\prc{Phys.~Rev.~C}%
\def\prd{Phys.~Rev.~D}%
\def\pre{Phys.~Rev.~E}%
\def\prl{Phys.~Rev.~Lett.}%
\def\pasp{PASP}%
\def\pasj{PASJ}%
\def\qjras{QJRAS}%
\def\rmxaa{Rev. Mexicana Astron. Astrofis.}%
\def\skytel{S\&T}%
\def\solphys{Sol.~Phys.}%
\def\sovast{Soviet~Ast.}%
\def\ssr{Space~Sci.~Rev.}%
\def\zap{ZAp}%
\def\nat{Nature}%
\def\iaucirc{IAU~Circ.}%
\def\aplett{Astrophys.~Lett.}%
\def\apspr{Astrophys.~Space~Phys.~Res.}%
\def\bain{Bull.~Astron.~Inst.~Netherlands}%
\def\fcp{Fund.~Cosmic~Phys.}%
\def\gca{Geochim.~Cosmochim.~Acta}%
\def\grl{Geophys.~Res.~Lett.}%
\def\jcp{J.~Chem.~Phys.}%
\def\jgr{J.~Geophys.~Res.}%
\def\jqsrt{J.~Quant.~Spec.~Radiat.~Transf.}%
\def\memsai{Mem.~Soc.~Astron.~Italiana}%
\def\nphysa{Nucl.~Phys.~A}%
\def\physrep{Phys.~Rep.}%
\def\physscr{Phys.~Scr}%
\def\planss{Planet.~Space~Sci.}%
\def\procspie{Proc.~SPIE}%
\let\astap=\aap
\let\apjlett=\apjl
\let\apjsupp=\apjs
\let\applopt=\ao
\bibliographystyle{mn2e}

\end{document}